\documentclass[11pt,twoside]{article}

\usepackage{asp2006}
\usepackage{natbib}
\usepackage{epsf}
\usepackage{lscape}

\begin{document}
\title{Increasing the Scientific Return of Stellar Orbits at the Galactic Center}   
\author{Sylvana Yelda\altaffilmark{1}, Andrea M. Ghez\altaffilmark{1}, 
Jessica R. Lu\altaffilmark{2}, Tuan Do\altaffilmark{1}, 
Will Clarkson\altaffilmark{1}, Keith Matthews\altaffilmark{2}}
\altaffiltext{1}{UCLA Department of Physics and Astronomy, Los Angeles, CA 90095-1547} 
\altaffiltext{2}{Astrophysics, California Institute of Technology, MC 249-17, 
Pasadena, CA 91125}

\begin{abstract} 
We report a factor of $\sim$3 improvement in Keck Laser Guide Star Adaptive
Optics (LGSAO) astrometric 
measurements of stars near the Galaxy's supermassive black hole (SMBH).
By carrying out a systematic study of M92, we have 
produced a more accurate model for the camera's optical
distortion. Updating our measurements with this model, and accounting for
differential atmospheric refraction, we obtain estimates of the SMBH properties
that are a factor of $\sim$2 more precise, and most notably, increase
the likelihood that the black hole is at rest with respect to the nuclear star
cluster.  These improvements have also allowed us to extend the radius 
to which stellar orbital parameter estimates are possible by a factor of 2.
\end{abstract}

\section{Introduction}
High angular resolution astrometry has been a very powerful technique for 
studies of the Galactic center (GC). Over the last decade, it has revealed a 
supermassive black hole \citep{eckart97, ghez98}, as well as a disk of young 
stars surrounding it \citep{levin03,genzel03,paumard06,lu09}.  
While the speckle imaging work carried out on the GC in the 1990's
had typical centroiding uncertainties of $\sim$1 mas, 
recent deep, adaptive optics (AO) images have improved the precision of stellar 
centroiding by a factor of $\sim$6-7, significantly increasing 
the scientific potential of astrometry at the GC
\citep{ghez08, gillessen09}. Further gains in astrometric precision 
could lead to ultra-precise measurements of the distance to the Galactic 
center (R$_o$), measurements of individual stellar orbits at larger 
galacto-centric radii, and, more ambitiously, to measure post-Newtonian 
effects in the orbits of short-period stars 
\citep[e.g.,][]{jarosz98,jarosz99,salim99,fragile00,rubilar01,weinberg05,
zucker07, kraniotis07,nucita07,will08}.

Two factors that currently limit astrometric measurements 
of stars at the GC are (1) the level to which AO cameras'
geometric distortions are known and (2) differential atmospheric 
refraction (DAR), which has not yet been explicitly corrected for in any
Galactic center proper motion study \citep{ghez08, gillessen09}.  While 
optical distortion from an infrared camera is expected to be static, 
distortion from the AO system and the atmosphere not corrected
by AO, is not. Initial estimates of the optical distortions for AO cameras
are generally based on either the optical design or laboratory test, which 
do not perfectly match the actual optical distortion of the system. 
Both uncorrected camera distortions and DAR leave $\sim$1-5 mas scale 
distortions over the spatial scales of the stars that are used to
define the reference frame for proper motions of stars
at the Galactic center.  It is therefore critical to correct these
effects.

In this contribution, we report our work to identify and correct for
two systematic errors, optical distortion and DAR (Section 2). Using
our updated astrometry, we obtain estimates of the central potential
that are a factor of 2 more precise than earlier work (Section 3), 
and we extend the radius to which stellar orbital parameter estimates 
are possible (Section 4).  A more detailed account of this work
will be reported in Yelda et al. (in preparation).

\section{Improving Keck/NIRC2 Distortion Models}
Observations of the globular cluster M92 (NGC 6341; 
$\alpha$ = 17 17 07.27, $\delta$ = +43 08 11.5) were made from
2007 June to 2009 May using the AO
system on the W. M. Keck II 10 m telescope with the facility 
near-infrared camera NIRC2 (PI: K. Matthews).  
M92 was observed at 79 different combinations of position angles (PAs) and
offsets, producing a final list of 2398 stellar 
positions.  These were compared to measurements obtained with ACS Wide Field 
Channel \citep[WFC; $\delta_{pos}\sim$0.5 mas,][]{anderson05} to derive a new 
NIRC2 distortion model, which produces a factor of 4 times smaller positional 
residuals than the previously used distortion
solution\footnote{http://www2.keck.hawaii.edu/inst/nirc2/preship\_testing.pdf}.

In addition to the correction for optical distortion, we implement for the
first time on Galactic center data a correction for differential atmospheric
refraction. DAR causes the separation between
a pair of stars viewed through the Earth's atmosphere to differ
depending on the zenith angle at which they are observed.  This effect can
be as large as $\sim$5 mas across our 10" field of view for the elevations at 
which the Galactic
center is observed at Keck.  We therefore correct for this effect based
on the prescriptions described in \citet{gublerTytler98}.

\section{Improved Galactic Center Astrometry}
With these changes, the Galactic center astrometry (absolute and relative) 
at Keck has significantly improved over previous work 
\citep[e.g.,][]{ghez08,lu09}. For the case of absolute astrometry, the
position and velocity of Sgr A*-radio in the IR reference frame improved
by a factor of $\sim$2 when computed in the same way as in \citet{ghez08}, who
used the overly-conservative approach of a half-sample bootstrap. Using 
the more appropriate jack-knife analysis, the errors improved by an additional
factor of $\sim$3 to $\sim$0.6 mas and $\sim$0.2 mas/yr for the position and velocity 
errors, respectively. For the case of relative astrometry, our uncertainties in the
position, velocity, and acceleration for stars brighter than K=15.5 are now
0.1 mas ($\sim$1 AU), 0.04 mas/yr ($\sim$1.5 km/s), 
and 0.02 mas/yr$^2$ ($\sim$0.8 km/s/yr), respectively.  With acceleration 
errors that are a factor of 3 times better, we are now constraining
orbital parameters for stars at distances of R$\sim$2" (two times further
than our earlier work).  
\begin{figure}[!ht]
\plottwo{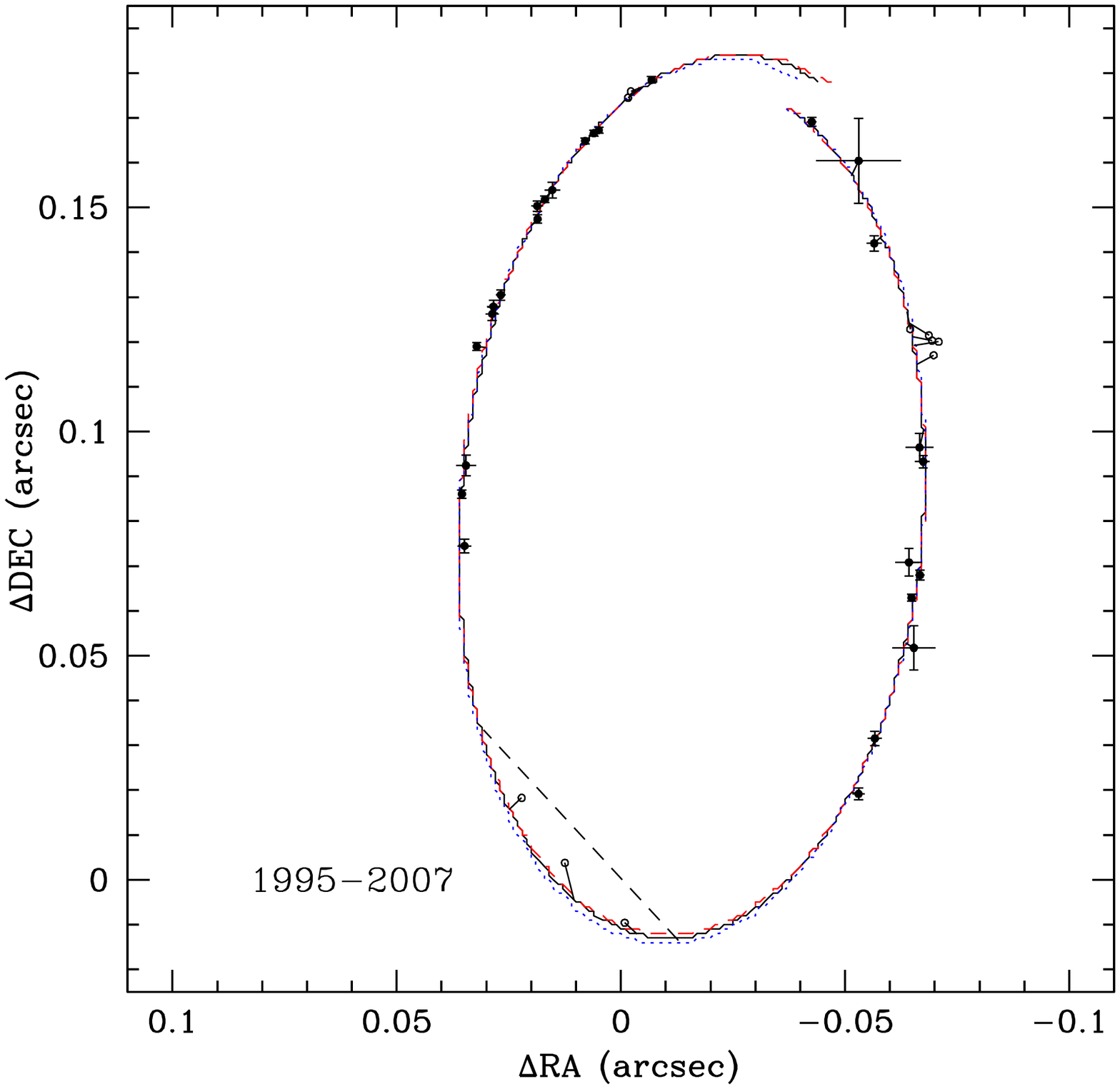}{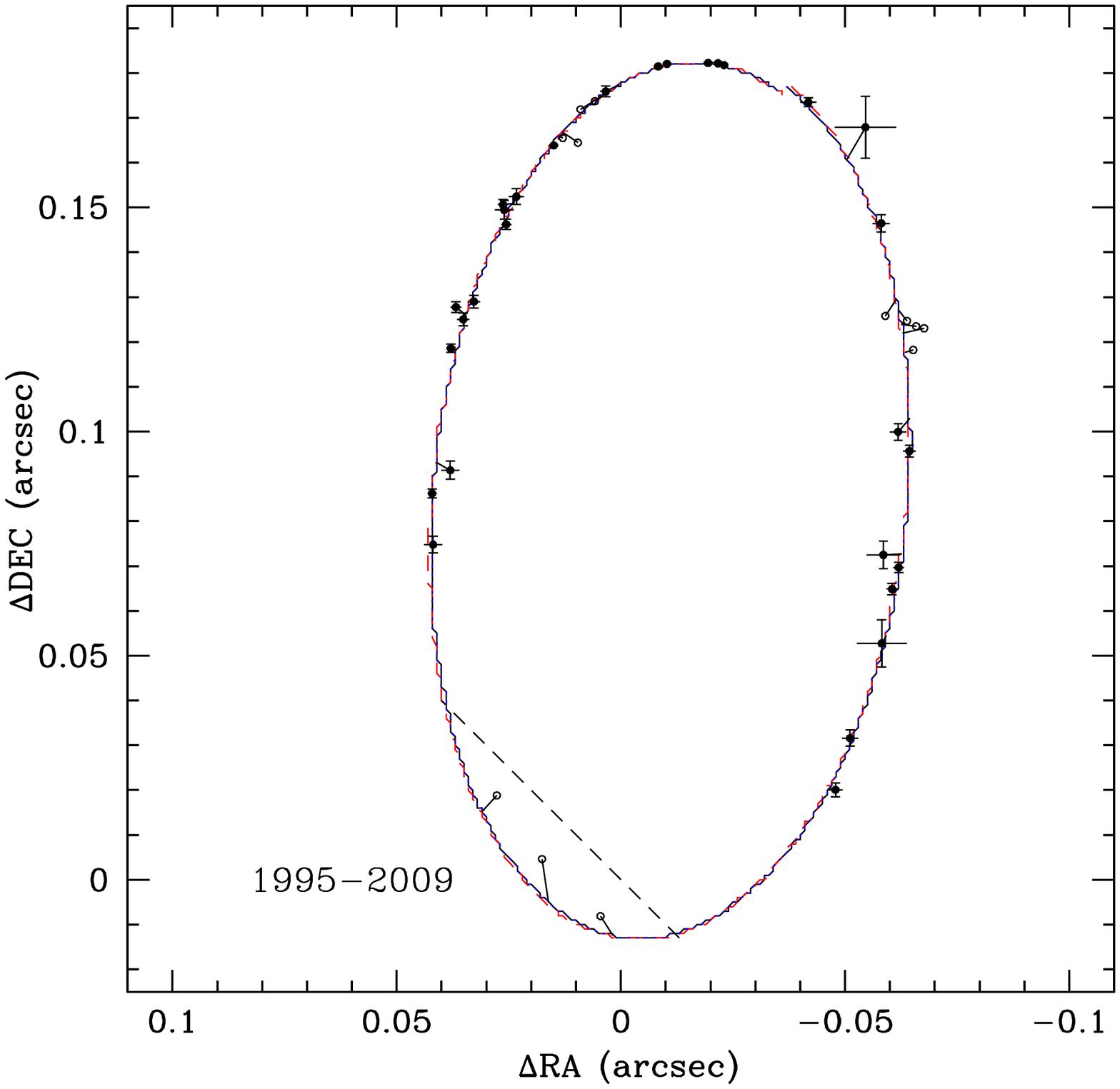}
\caption{{\em Left:} Orbit from Ghez et al. (2008) showing their best fit
to the astrometric data of S0-2 ({\em solid}). 
{\em Right:} Best fit orbit after implementing the
new distortion solution and DAR correction ({\em solid}). 
The dotted (blue) and dashed (red) orbits show additional solutions which give 
the minimum and maximum velocity, respectively, for Sgr A* relative to the cluster
for solutions with ($\chi^2 - \chi^2_{min}$) $< $ 1.
Note that all orbits assume no priors on the black hole's position or motion.  
}
\end{figure}

In recent orbital analyses, the need to introduce a term to account
for possible relative motion between the black hole and the reference frame has
been emphasized as a way to account for systematic errors that might arise
from instabilities in the reference frame \citep{ghez08,gillessen09}.
Using our improved astrometry from measurements reported in \citet{ghez08}
as well as new measurements taken in 2008 and 2009, we obtain revised 
estimates for the orbit of S0-2, which are compared
with that obtained in \citet{ghez08} in Figure 1. Two additional orbital solutions are
plotted over the best fit solutions; these show solutions giving the smallest
and largest velocities for Sgr A* for solutions with ($\chi^2 - \chi^2_{min}$) $< $ 1.
The new orbital analysis allows for a smaller velocity for Sgr A*, increasing
the likelihood that the black hole is at rest relative to the nuclear stellar
cluster.  We also find similar values for the black hole mass and the distance
to the GC as in our earlier work, but with smaller errors.  
Figure 2 shows a comparison of the resulting fit parameters 
before and after this work. We find for the black hole 
mass $M_{BH}$=4.1$\pm$0.3 $M_\odot$ and for the distance to the Galactic
center $R_0$=8.0$\pm$0.3 kpc.  
This exercise was repeated using only the data sets used in \citet{ghez08}
and we find similar values for the fit parameters, within the uncertainties.

\begin{figure}[!ht]
\plottwo{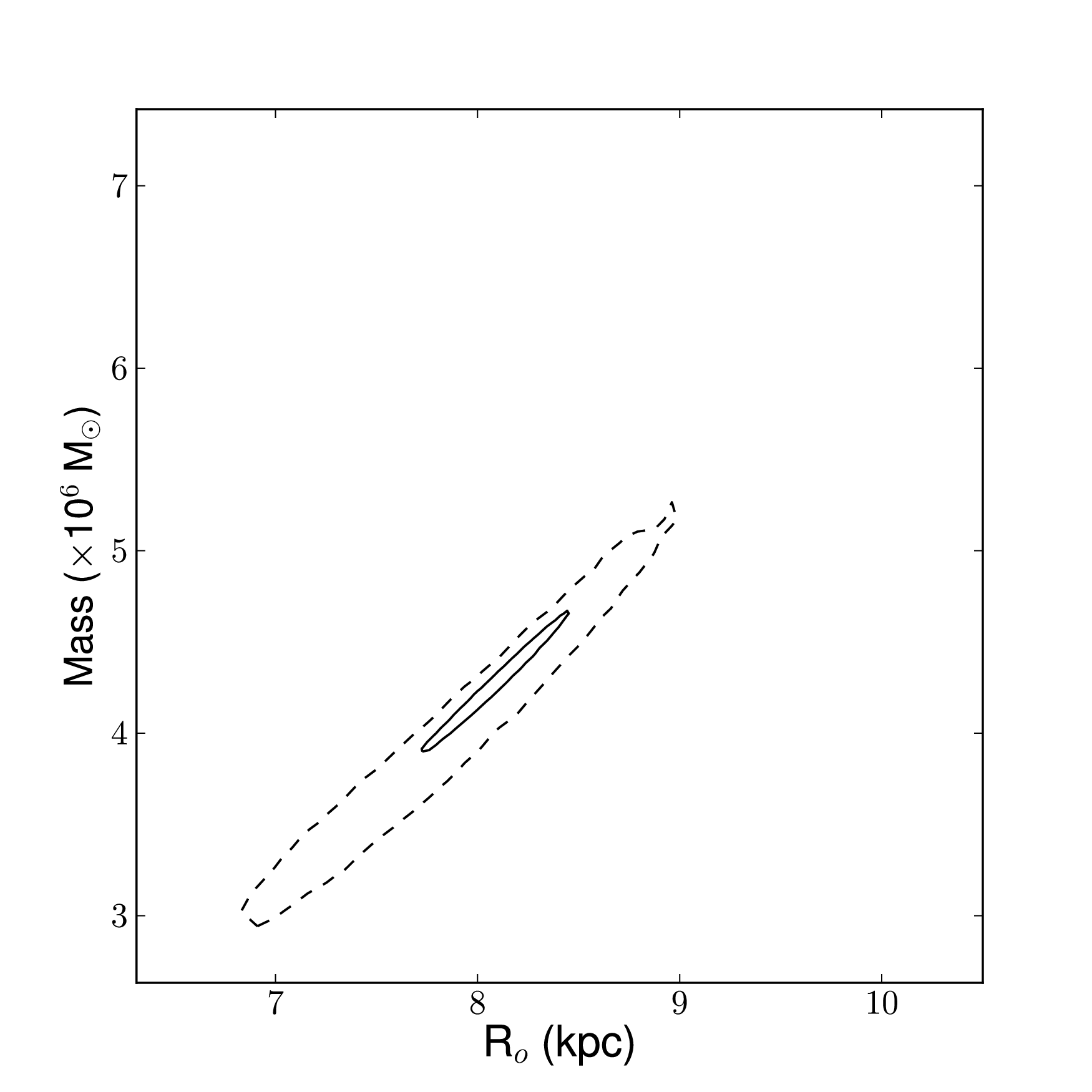}{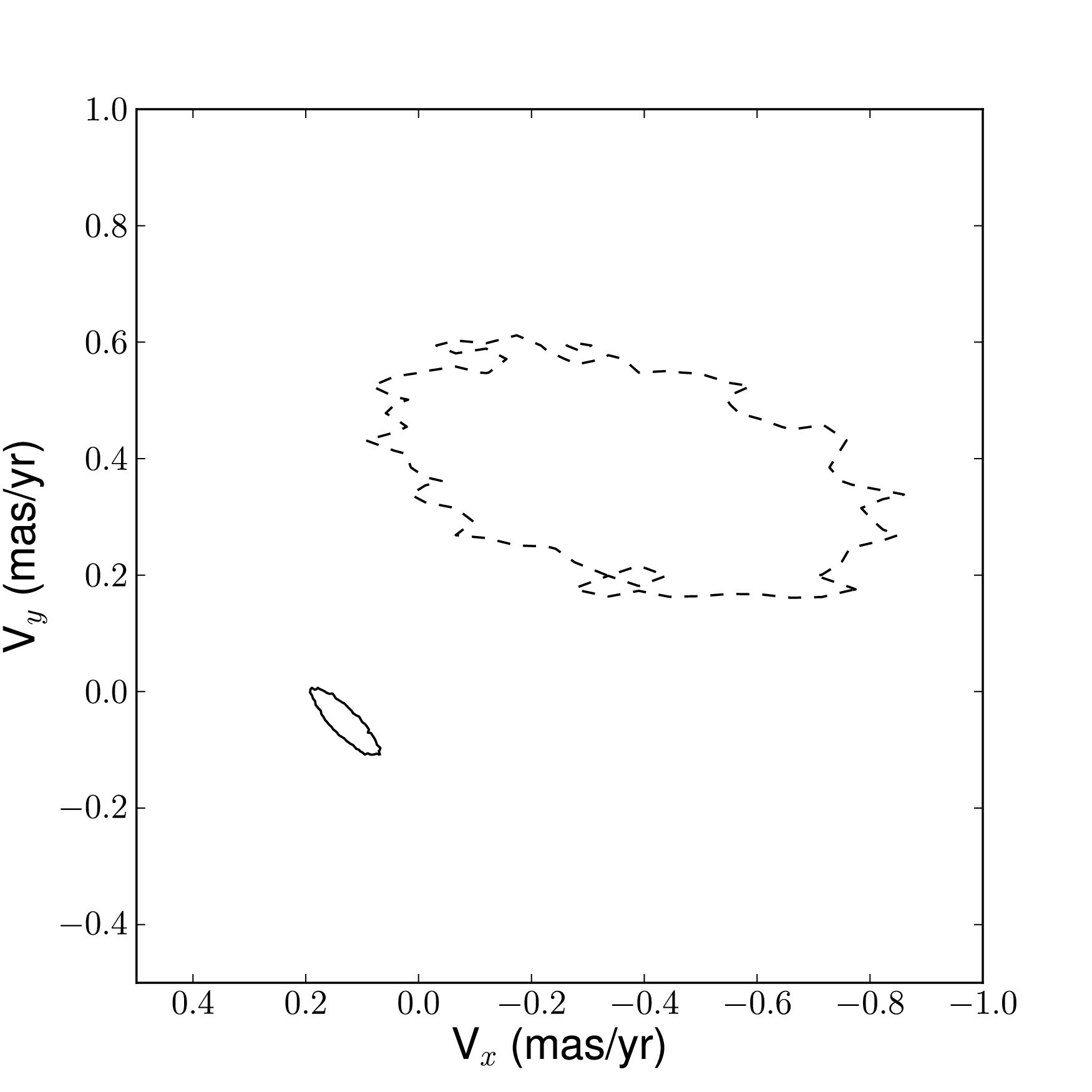}
\caption{One sigma contours from the orbit of 
S0-2 showing the fit parameters of black hole mass versus distance to the GC 
({\em left}) and velocity of the black hole in the plane of the sky ({\em right}). 
The dashed contours are from Ghez et al. (2008), while the solid contours show 
results from this work.
The improved astrometric reference frame has resulted in tighter 
constraints on the properties of the black hole.}
\end{figure}

\section{Conclusions}
We have improved upon existing geometric distortion solutions for the 
NIRC2 camera at the W. M. Keck II telescope and have, for the first 
time, implemented DAR corrections to our Galactic center astrometry.  
With our improved astrometric reference frame we find a nearly closed
orbit for the short-period star, S0-2, with little motion allowed for
the central black hole. Furthermore, we have detected accelerations
in the plane of the sky out to a projected distance of R$\sim$2".

\acknowledgements 
This work was supported by NSF grant AST 09-09218.
The W.M. Keck Observatory is operated as a scientific partnership among 
the University of California, the California Institute of Technology 
and NASA. The Observatory was made possible by the generous financial 
support of the W.M. Keck Foundation.

\end{document}